\documentclass[journal]{IEEEtran}

\ifCLASSINFOpdf
\else
   \usepackage[dvips]{graphicx}
\fi
\usepackage{url}

\usepackage{amsmath,graphicx}
\usepackage{adjustbox}
\usepackage{subcaption}

\begin{document}

\title{Interfacing PDM MEMS microphones with PFM spiking systems:  Application for Neuromorphic Auditory Sensors}

\author{A. Jimenez-Fernandez, D. Gutierrez-Galan, A. Rios-Navarro, J. P. Dominguez-Morales and G. Jimenez-Moreno
\thanks{This work has been submitted to the IEEE for possible publication. Copyright may be transferred without notice, after which this version may no longer be accessible.

This work was supported by the Spanish grant (with support from the European Regional Development Fund) COFNET (TEC2016-77785-P). The work of Daniel Gutierrez-Galan was supported by a Formaci\'{o}n de Personal Investigador Scholarship from the Spanish Ministry of Education, Culture and Sport.}
\thanks{All authors are with the Universidad de Sevilla,ETS Ingenieria Informatica. Avd. Reina Mercedes s/n, Sevilla, Spain (e-mail: angel@us.es).}
}

\markboth{Journal of \LaTeX\ Class Files, Vol. 14, No. 8, August 2015}
{Shell \MakeLowercase{\textit{et al.}}: Bare Demo of IEEEtran.cls for IEEE Journals}
\maketitle

\begin{abstract}
In neuromorphic engineering, computation is commonly performed asynchronously, mimicking the way in which nervous systems process information: spike by spike. The Neuromorphic Auditory Sensor (NAS) has been implemented under this principle: applying different spike-based Signal Processing blocks.
Computation in the spike domain requires the conversion of signals from analog or digital representation to the spike domain, which could present a speed constraint in many cases.  This paper presents a spike-based system to convert audio information from low-power pulse density modulation (PDM) MicroElectroMechanical Systems (MEMS) microphones into rate coded spike frequencies. These spikes represent the input signal of the NAS, avoiding the analog or digital to spike conversion, and therefore improving the time response of the NAS. This conversion has been done in VHDL as an interface for PDM microphones, converting their pulses into temporal distributed spikes following a pulse frequency modulation (PFM) scheme with an accurate Inter-Spike-Interval, known as "PDM to spikes interface" (PSI). This was tested in two scenarios, first as a stand-alone circuit for its characterization, and then integrated with a NAS for verification. The PSI achieves a Total Harmonic Distortion (THD) of -39.51dB and a Signal-to-Noise Ratio (SNR) of 59.12dB, demands less than 1\% of the resources of a Spartan-6 FPGA and has a power consumption below 5mW.
\end{abstract}


\IEEEpeerreviewmaketitle

\section{Introduction}
\label{sec:intro}

Pulse-density modulation (PDM) is a sigma-delta modulation technique used to digitize an analog signal with a 1-bit data stream and a high sample rate. In recent years, many low-power microelectromechanical (MEMS) microphones designed for mobile applications, such as tablets, laptops and cell phones, among others, have appeared in the market. In PDM data streams, a logic ‘1’ corresponds to a pulse of the maximum positive polarity (+A), and a logic ‘0’ represents the maximum negative polarity (-A). A signal value of 0 is codified by an alternation of ‘1’s and ‘0’s. Commonly, this type of modulation is associated with neuromorphic information codification, in the sense of being a rate-coded signal \cite{smith2010neuromorphic}. This kind of computation allows processing information only when it is needed, avoiding periodic or redundant data processing, thus saving power and computational resources \cite{liu2019}.

Currently, we can find diverse neuromorphic cochleae, both analog \cite{chan2007aer}\cite{ hamilton2008}\cite{ wen2009}\cite{liu2010} and digital  \cite{Mugliette2011, thakur2014}, inspired by Lyon’s cascade model \cite{lyon82} modeling the inner-hair cells (IHC). In \cite{jimenez2017binaural}, a Neuromorphic Auditory Sensor (NAS) is presented, based on spike signals processing (SSP) techniques \cite{jimenez2010building}\cite{jimenez2012neuro}.

\begin{figure*}[t]
\centering\includegraphics[width=0.75\linewidth]{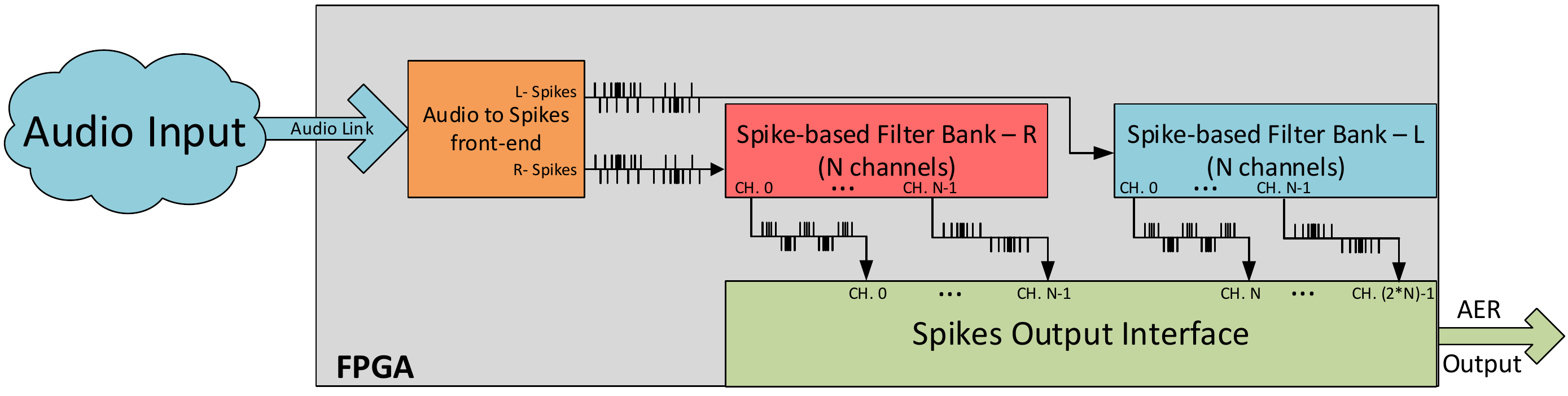}
\caption{NAS Architecture: Audio to spikes, spikes processing banks, and AER output interface.}
\label{fig_Global_NAS_Arch}
\end{figure*}

 Fig. \ref{fig_Global_NAS_Arch} shows a global scheme of the NAS architecture. First, the audio information is provided by a digital audio codec, whose discrete audio samples output is converted into spike streams, following the pulse frequency modulation (PFM). The NAS filters these spikes directly, spike after spike, using a set of Spike-based Low-Pass Filters (SLPF) connected in a cascade fashion. Finally, spikes are transmitted to the next layers using the Address-Event Representation (AER) protocol \cite{boahen2000point}. 

NAS has been currently used for many practical applications, as pitch frequency detection \cite{dominguez2016multilayer}, musical tones identification \cite{cerezuela2016sound}, sound source localization \cite{schoepe2019neuromorphic}, heart murmurs diagnosis \cite{dominguez2018murmurs}, and speech recognition \cite{dominguez2018deep}, among others. Great effort has been dedicated to improve NAS features, as it is the input layer of all these systems, improving responses and spreading for new applications of this technology.

One main disadvantage of the NAS is the need for a discrete audio codec to capture analog audio. Audio codecs provide a set of digital periodic samples that must be converted into spikes. These devices have a sampling period from 22.67$\mu$s to 10.41$\mu$s, limiting the temporal capabilities, e.g., sound localization applications. \cite{indivieri19}However, PDM microphones provide a stream of rate-coded signals with higher sample rate (3.125MHz in this case, with a time resolution of 320ns), which can represent the NAS input and be directly processed  as spiking information. Therefore way, the need to generate spike streams synthetically is avoided, which was a restriction in previous NAS implementations \cite{jimenez2017binaural}.

\section{PDM to spikes interface (PSI)}
\label{sec:PDMtoSpikes}

PDM information codification is substantially different from rate-coded spike-based signals. In rate-coded spike-based signals, the information is given by the spikes frequency, which means that the information is inversely proportional to the temporal Inter-Spike-Interval (ISI). This means that, with only two spikes, it is possible to reconstruct the amplitude of the original signal. Spike-based systems use PFM  to distribute the spikes in time properly, in order to accurately represent the signal's information. In PDM signals, the information is contained in the density of pulses, and one pulse is generated every clock cycle, where a logic '1' represents a positive value, and a logic '0' a negative one. For example, when there are more ‘1’s than ‘0’s the information is positive, and the more ‘1’s, the higher the amplitude is. Thus, for reconstructing the signal's amplitude, it is necessary to collect PDM pulses during a temporal window, performing a downsampling operation.

Digital systems convert PDM signals to digital values using the pulse coded modulation (PCM). PCM is reconstructed from PDM with a digital decimation stage, commonly performing a downsampling by a factor of 64, and providing a multiple-bits word (e.g., 16 bits @ 48.8kSamples/s) with high frequency noise added. After this stage, an infinite impulse response (IIR) filter is commonly used as a band-pass filter (BPF) to remove DC components and high frequency quantization noise.

\begin{figure}[ht]
\centering\includegraphics[width=0.9\linewidth]{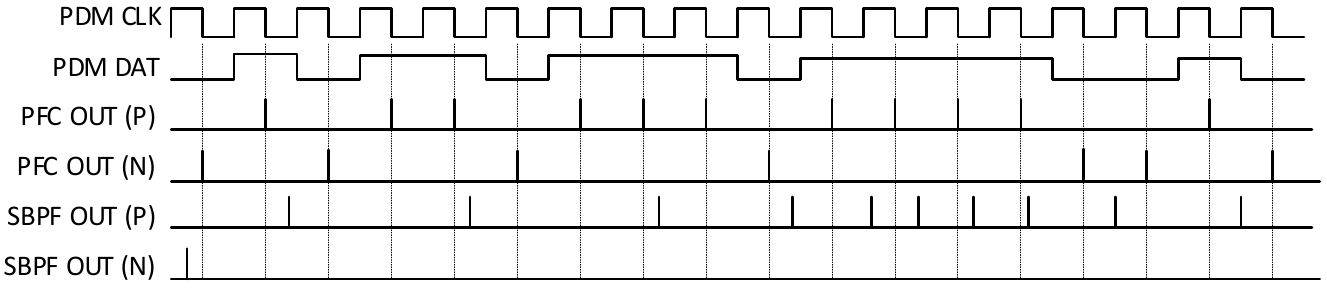}
\caption{Filtered spikes evolving from an increasing PDM audio signal.}
\label{fig_filtered_spikes}
\end{figure}

The main goal of this work is to design an HDL circuit able to read PDM pulses and redistribute them in time as rate-coded spikes, with an ISI proportional to the sound pressure. Fig. \ref{fig_filtered_spikes} briefly shows how signals evolve from PDM pulses to PFM spikes.

\begin{figure}[ht]
\centering\includegraphics[width=0.9\linewidth]{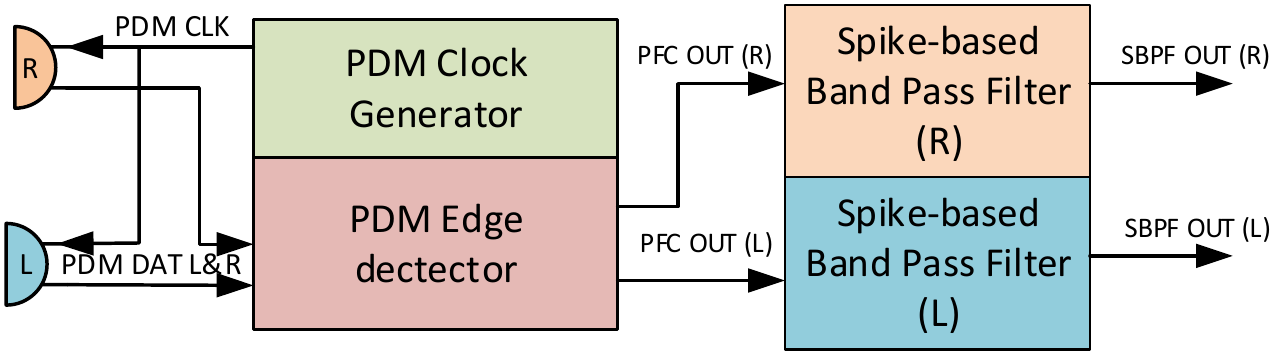}
\caption{PDM to spikes interface circuit.}
\label{fig_PDM_to_spikes_interface}
\end{figure}

To convert PDM information into rate-coded spikes, a two stages circuit (Fig. \ref{fig_PDM_to_spikes_interface}) is proposed. The first stage is a finite state machine (FSM) circuit that works as an edge detector, and generates a spike of a single clock cycle for each PDM pulse. The next stage consists in one (monaural) or two (binaural) banks of spike-based band-pass filters (SBPF), which process raw spikes from the FSM to give a temporal distributed spikes stream.

Since spikes can be both positive and negative, we use two wires to represent signed spikes. The FSM output generates a stream of signed spikes that are still not distributed in time, with the ISI being constant and equal to the PDM clock period. Fig. \ref{fig_filtered_spikes} presents an example of a positive increasing audio signal, and how spikes evolve.

\subsection{PDM front-end circuit}
The PDM front-end circuit (PFC) has two main functionalities: to generate the PDM clock and to convert long PDM pulses into one clock cycle spikes. The hardware platform used to implement these blocks is called AER-Node \cite{Iakymchuk2014} and it has a clock frequency of 50MHz. Dividing this clock by a factor of 16, we get a PDM clock of 3.125MHz, which is the maximum value allowed by this kind of MEMS microphones. In every PDM clock cycle there is a PDM pulse in the PDM DAT line. If PDM DAT has a value of ‘1’ then a positive spike is transmitted to the next stage, and if there is a ‘0’ it will be a negative spike.

\subsection{Second-order Spikes Band-Pass Filter (SBPF)}
\begin{figure}[th]
\centering\includegraphics[width=0.85\linewidth]{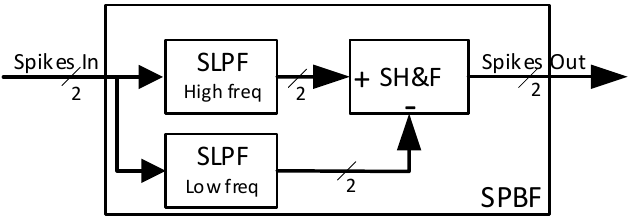}
\caption{Spike Band-Pass Filter (SBPF) internal blocks.}
\label{fig_SBPF}
\end{figure}

The next stage is a Spike Band-Pass Filter (SBPF), whose functionality is detailed in \cite{dominguez2011designing}. This filter is composed of two first-order Spike-based Low-pass filters (SLPF) and one Spike Hold \& Fire (SH\&F) (see Fig. \ref{fig_SBPF}). SH\&F is a SSP building block that subtracts the spike rate between two spiking signals (detailed in \cite{jimenez2012neuro}). The SLPF that is connected to the SH\&F's positive input has a cut-off frequency that is higher than the SLPF connected to the negative input. Subtracting the output from both spike-based filters, only the information in the middle band remains, rejecting the DC and high-frequency components. These filters are connected with 2-bit buses (for positive and negative spikes). These blocks use positive and negative activity to represent the bipolar nature of audio.

%Equation (1) is used for modeling the SBPF behavior. In this equation, both filters transfer functions are subtracted, getting a new equivalent transfer function, where k\textsubscript{SLPF} is the SLPF gain, and w\textsubscript{SLPF} is the cut-off frequency, for higher and lower frequency filters respectively.

%If we set both SLPFs gain to 1 (0dB), this equation is simplified, and is similar to a 2º order band pass filter with mid-frequency (w\textsubscript{MID}), quality factor (Q), and band pass gain (k\textsubscript{SBPF}) dependent on the SLPF cut-off frequencies, as (2), (3) and (4) describe.

%\begin{equation}
%\begin{aligned}
 %   F_{SBPF}\, (s)=F_{SLPF_{HF}}(s)-F_{SLPF_{LF}}(s)= \\ \frac{k_{spikesDivHF}\:  *\:  k_{BWSpikesGenHF}}{s + k_{spikesDivFBHF}\:  *\:  k_{BWSpikesGenHF}}\\ - \frac{k_{spikesDivLF}\:  *\:  k_{BWSpikesGenLF}}{s + k_{spikesDivFBLF}\:  *\:  k_{BWSpikesGenLF}}
%\end{aligned}
%\end{equation}

%\begin{equation}
%\begin{aligned}
 %   \omega_{mid} = \sqrt{\omega_{HF}*\omega_{LF}}
%\end{aligned}
%\end{equation}

%\begin{equation}
%\begin{aligned}
%    Q = \frac{\sqrt{\omega_{HF}*\omega_{LF}}}{\omega_{HF}+\omega_{LF}}
%\end{aligned}
%\end{equation}

%\begin{equation}
%\begin{aligned}
%    k_{BPF} = \frac{\omega_{HF}-\omega_{LF}}{\omega_{HF}+\omega_{LF}}
%\end{aligned}
%\end{equation}

\subsection{Hardware resources and power consumption}
The PSI design was synthesized and implemented on a Xilinx Spartan 6 FPGA (XC6LX150T) to measure the required resources and its power consumption. Table \ref{table_HardwareReq} presents the resources that are needed to implement PSI in FPGA. As can be seen, the amount of resources needed is under 0.45\% of the total slice registers and logic (LUT) of the FPGA. The PSI can operate at a clock frequency up to 147.18 MHz. After the synthesis, the power consumption was simulated using Xilinx XPower tool assuming a 50\% of signal transitions, obtaining a power consumption estimation of 2.67mW for the PSI. This power consumption should be added to the MEMS microphones' power, which depends on the ones that are used. In our case, each of the microphones demands 0.98mW (according to the documentation provided by the manufacturer). Therefore, the whole system demands a total of 4.63mW for a binaural NAS.

%\begin{adjustbox}{max width=\textwidth}

\begin{table}[ht]

\caption{PSI hardware requirements}
\label{table_HardwareReq}
\centering
\begin{adjustbox}{max width=0.47\textwidth}
\begin{tabular}{|c|c|c|}
\hline
\multicolumn{3}{|c|}{\textbf{\begin{tabular}[c]{@{}c@{}}Post-implementation results (Spartan 6 - XC6SLX150T)\end{tabular}}} \\ \hline
\textit{\textbf{Slices Registers (\%)}}      & \textit{\textbf{Slices LUT (\%)}}      & \textit{\textbf{Max Clock Freq.}}     \\ \hline
204 / 184.304 (0.11\%)                          & 409 / 92.152 (0.44\%)                     & 147.18 MHz                            \\ \hline
\end{tabular}
\end{adjustbox}
\end{table}

\section{Experimental setup}
\label{sec:pagestyle}

\begin{figure}[h]
\centering\includegraphics[width=0.9\linewidth]{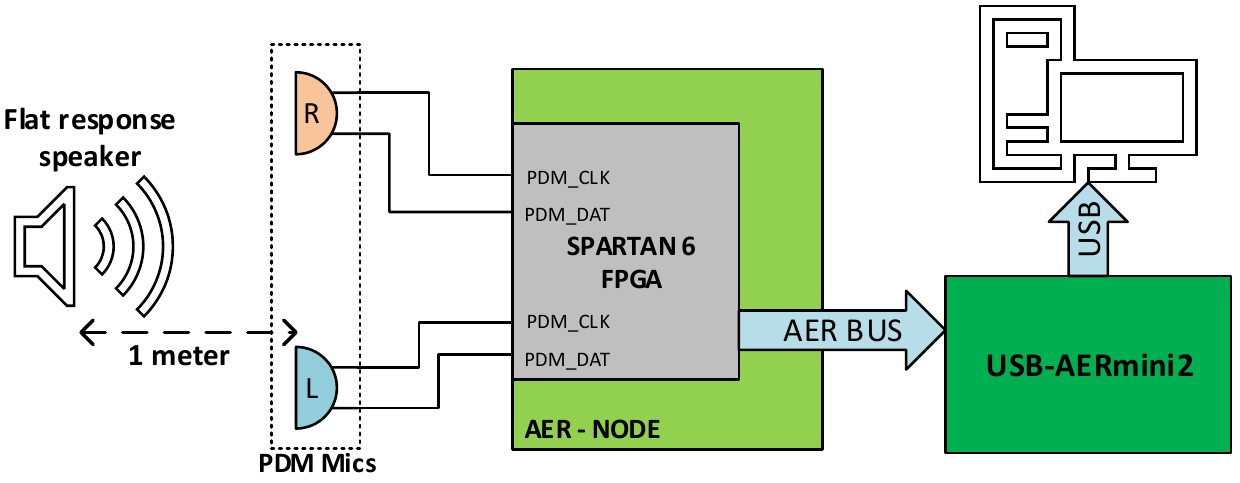}
\caption{Test scenario. Sound is played by a  response speaker, exciting PDM microphones. Finally, the information is sent to a computer through an AER-to-USB interface.}
\label{fig_experimental_results}
\end{figure}

For testing purposes, a scenario was built to analyze the PSI's standalone behavior. Fig. \ref{fig_experimental_results} presents the testing setup, where two PDM microphones from ST Microelectronics (MP34DT02) were connected to an AER-Node board, which was in turn connected to an USB-AERmini2 board. MP34DT02 are omnidirectional MEMS microphones with PDM interfaces, with an acoustic overload point of 120dB\textsubscript{SPL}, an SNR of 60dBm, a dynamic range of 86dB, and a maximum power consumption of 0.98mW (as previously described). 

The AER-Node board has a Xilinx Spartan 6 FPGA (XC6S150T), which holds the PSI, a 128-channel binaural NAS, and a set of AER interfaces. Its parallel AER output was connected to the USB-AERmini2 board \cite{berner2007}, which works like a bridge between AER buses and USB ports, allowing the AER events to be sent from the AER-Node board to a host computer. In the computer, two software tools were running: jAER \cite{delbruck2008frame}, to visualize and log AER information; and MATLAB, to analyze and process the events. The sound used to excite the system was played using a flat response audio speaker, in this case a BEHRITONE C5A from Behringer, placed at a 1-meter distance from the PDM microphones and at a fixed gain in order to have an audio level of 65dBSPL on the microphones' side. This kind of equipment was used to avoid the influence of audio equalizers and the compensation that domestic Hi-Fi equipment presents. Thus, no preprocessed sounds were used and, instead, we tried to reproduce sound waves in the most ideal way possible. This will potentially open our system to many stand-alone applications, such as robotics.

\subsection{PSI Experimental results}
\label{sec:typestyle}

\begin{figure}[ht]
\centering\includegraphics[width=0.8\linewidth]{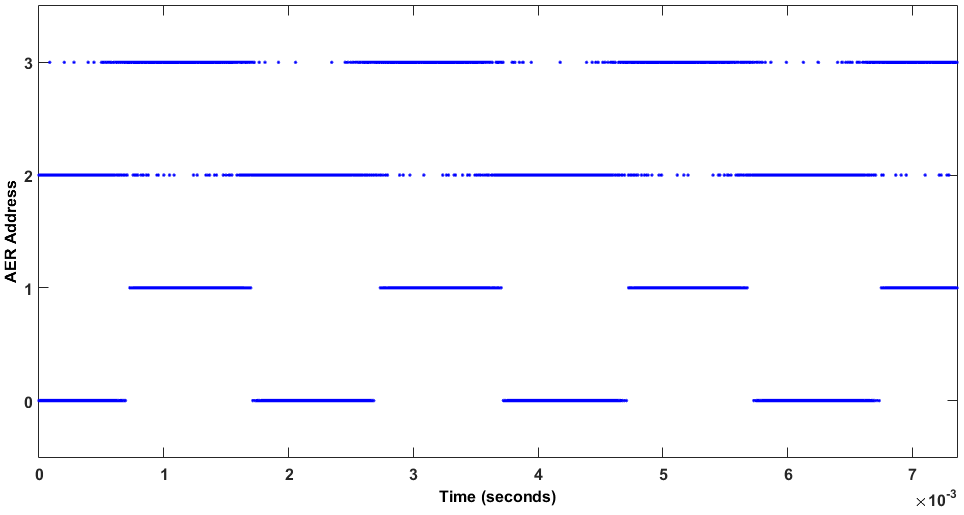}
\caption{Spikes from PSI: PDM front-end output (3-2) (top), and PSI's output (1-0) after filtering (bottom).}
\label{fig_spikes_from_PSI}
\end{figure}

For the first experiment, the system was stimulated with a clear 500Hz pure tone audio signal played by the flat response speaker. Fig. \ref{fig_spikes_from_PSI} represents the spikes from each stage of the PSI. Higher addresses (3 and 2) correspond to the spikes fired by the PDM front-end circuit, and lower addresses (1 and 0) to the SPBF output. Spike addresses 3 and 1 are positive, whilst 2 and 0 are negative.

Fig. \ref{fig_spikes_from_PSI}  depicts how the addresses that contain the output of the PDM front-end overlap the information between positive and negative, which does not happen after filtering it with the PSI. In PDM, information makes sense for the average activity of a temporal window. However, in the spikes domain, the information is coded with the time between two consecutive spikes. From the signal sign point of view, we can say that zero-crossing is performed when the polarity of the spikes changes(i.e. after a positive spike, a negative one is produced). In the case of the PDM front-end output, there are several spikes overlapping positive (address 3) and negative activity (address 2). From the point of view of ISI, this represents a considerable amount of high-frequency noise. However, if we check the SBPF output spikes, there is no overlapping between positive (address 1) and negative (address 0) activity, rejecting high frequency noise.

\begin{figure}[ht]
\centering\includegraphics[width=0.9\linewidth]{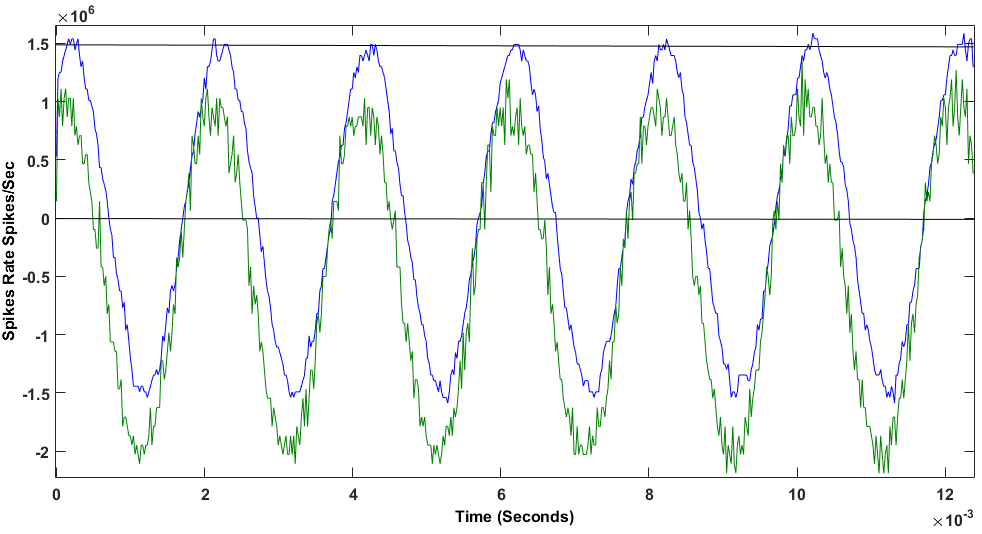}
\caption{Temporal reconstruction of a 500Hz tone. Green: PDM front-end's output. Blue: SBPF's output.}
\label{fig_temporal}
\end{figure}

\begin{figure}[ht]
\centering\includegraphics[width=1.0\linewidth]{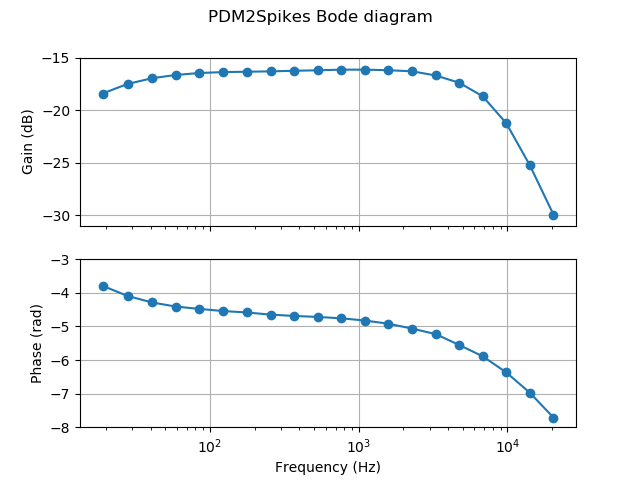}
\caption{Bode diagram of the PDM2Spikes module.}
\label{fig_bode}
\end{figure}

Fig. \ref{fig_temporal} shows the reconstruction of the original signal using the spikes' ISI. First, the green signal represents the reconstruction from PDM front-end's output. This is a noisy signal and it has an offset introduced by the PDM microphones. On the other hand, the blue signal is the reconstruction from SBPF’s output. A clear tone with neither noise nor offset can be seen, improving the previous audio signal quality. Analyzing this response, we achieve a Total Harmonic Distortion (THD) of -39.51dB and a Signal-to-Noise Ratio (SNR) of 59.12dB.

To measure the number of zero-crossings, a one second recording was analyzed and the amount of changes from positive spike to negative and vice versa were counted. In the PDM front-end's output, more than 80k zero-crossings were found. However, in SBPF's output, 1k zero-crossings were found, which exactly matches a 500Hz signal.

Our second experiment consisted in a frequency sweep from 20Hz to 20KHz to analyze the behavior of the system with different frequencies. Fig. \ref{fig_bode} shows the frequency sweep results as a bode diagram. The top curve in Fig. \ref{fig_bode} presents the gain for diverse frequencies. PSI gain starts to increase from 70Hz to 12KHz, and then decreases rejecting higher frequencies. This bandwidth is enough for many applications related to speech and speakers recognition. The spike-based filters in the PSI introduce a temporal deviation. It was measured as signal phase (in rads) and the results are included in Fig. \ref{fig_bode} bottom. PSI has a mean phase of -4.5 rads, approximately, increasing when frequency is close to the cut-off frequency, as expected from a low-pass filter.

\subsection{NAS integration}
\label{sec:majhead}

\begin{figure}
    \centering
    \begin{subfigure}[ht]{0.48\textwidth}
        \includegraphics[width=\textwidth]{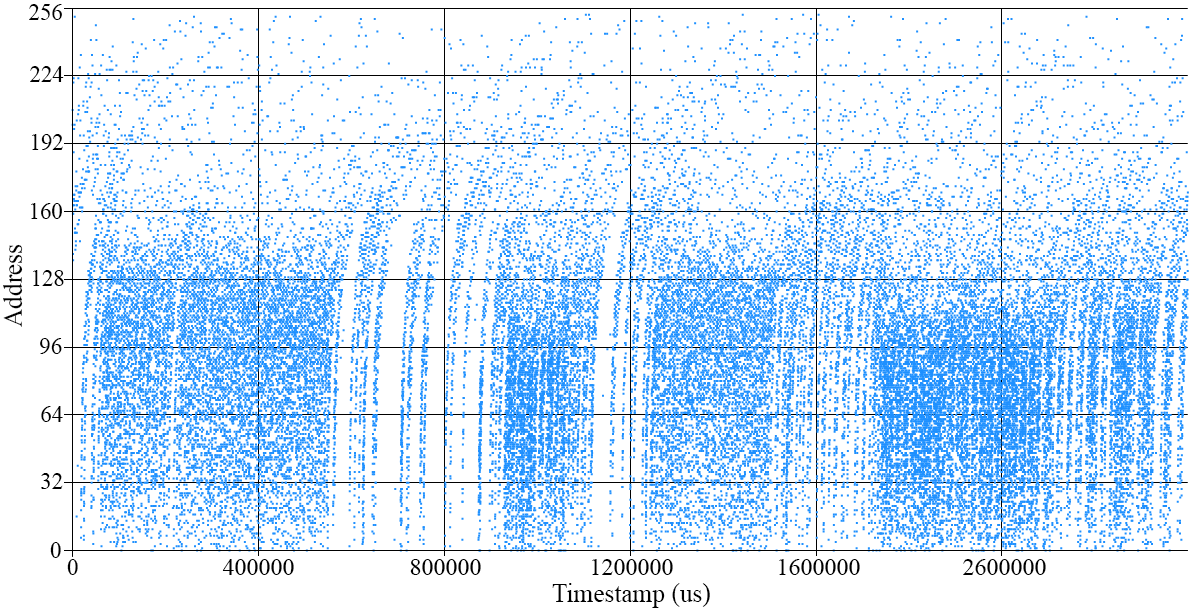}
    \end{subfigure}
    ~ %add desired spacing between images, e. g. ~, \quad, \qquad, \hfill etc. 
    \begin{subfigure}[ht]{0.48\textwidth}
        \includegraphics[width=\textwidth]{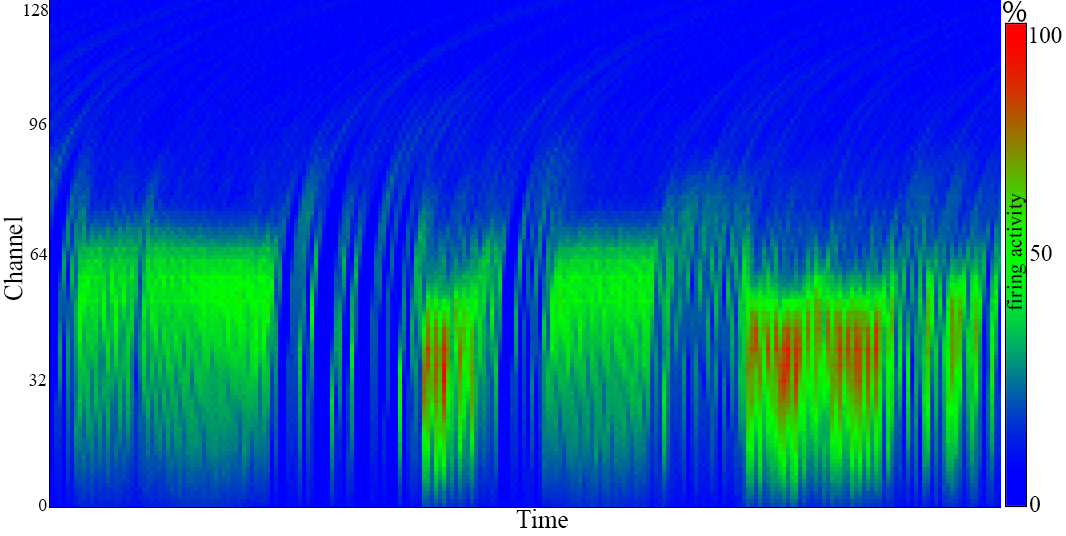}
    \end{subfigure}
    ~ %add desired spacing between images, e. g. ~, \quad, \qquad, \hfill etc. 
    \caption{Cochleogram (top) and sonogram (bottom) obtained with NAVIS from a speaker saying ``Si vis pacem para bellum''.}\label{fig:fig_si_vis_coc_sonog}
\end{figure}

In order to validate the PSI on a real scenario, it was integrated in a 128-channel binaural NAS. This NAS was fed with a male voice saying: ``Si vis pacem, para bellum'', and the output activity was recorded using an USB-AERMini2 board as an AER-DATA file. Fig.\ref{fig:fig_si_vis_coc_sonog} contains the cochleogram and the sonogram of this recording, respectively. Each word is clearly distinguishable, and activates middle channels between 200Hz and 5kHz. These figures were obtained by using NAVIS software \cite{dominguez2017navis}.

%\begin{figure}[ht]
%\centering\includegraphics[width=0.85\linewidth]{Images/coc.png}
%\caption{NAVIS cochleogram: ``Si vis pacem para bellum''.}
%\label{fig_si_vis_coc}
%\end{figure}

%\begin{figure}[ht]
%\centering\includegraphics[width=0.85\linewidth]{Images/sonogram.png}
%\caption{NAVIS sonogram: ``Si vis pacem para bellum''.}
%\label{fig_si_vis_sonog}
%\end{figure}

\section{Conclusions}
\label{sec:print}

In this paper, a PDM to PFM Spikes circuit is presented. PDM MEMS microphones are useful for low-power, stand-alone, embedded applications. Their output is based on spike density, and it needs to be adapted in order to be used as input to the NAS. A two-stage circuit for FPGA was designed, which is able to convert PDM information to PFM spikes with a consistent ISI. The PSI was synthesized for a Spartan 6 FPGA with low resources and power requirements. It was then tested with real audio stimulus, analyzing its behavior in terms of temporal response and zero-crossings. The PSI was also integrated in a full NAS to demonstrate the viability of the combination of this kind of systems. The use of PDM microphones with NAS considerably simplifies the system, enabling compact and portable spike-based auditory systems with lower power consumption.

\bibliographystyle{IEEEtran}
\bibliography{references.bib}

\end{document}